\documentclass[
aps,
pra,
twocolumn,
superscriptaddress,
floatfix,
longbibliography
]{revtex4-1}

\usepackage[dvipdfmx]{graphicx}
\usepackage{dcolumn}
\usepackage{ulem}

\usepackage{bm}
\usepackage{mathtools}
\usepackage{braket}
\usepackage{color}

\usepackage{physics}

\usepackage{amsmath}
\usepackage{amssymb}
\usepackage{amsthm}

\usepackage{commath}
\usepackage{autobreak}

\usepackage{enumitem}

\theoremstyle{definition}

\newtheorem*{theorem*}{Theorem}

\newcommand{\pair}[2]{\langle #1, #2\rangle}
\usepackage[colorlinks=true]{hyperref}

\usepackage{CJKutf8}

\begin{document}

\title{Emergent long-range interaction and state-selective localization in a strongly driven $ XXZ $ model}

\author{Kentaro Sugimoto}
 \email{tarotene@iis.u-tokyo.ac.jp}
 \affiliation{Department of Physics, the University of Tokyo, 
 Kashiwa, Chiba 277-8574, Japan}
 \affiliation{Computational Condensed Matter Physics Laboratory,
 RIKEN Cluster for Pioneering Research (CPR), Wako, Saitama 351-0198, Japan}
\author{Seiji Yunoki}
 \affiliation{Computational Condensed Matter Physics Laboratory,
 RIKEN Cluster for Pioneering Research (CPR), Wako, Saitama 351-0198, Japan}
 \affiliation{Computational Quantum Matter Research Team,
 RIKEN Center for Emergent Matter Science (CEMS), Wako, Saitama 351-0198, Japan}
 \affiliation{Computational Materials Science Research Team,
 RIKEN Center for Computational Science (R-CCS), Kobe, Hyogo 650-0047, Japan}
 \affiliation{Quantum Computational Science Research Team,
 RIKEN Center for Quantum Computing (RQC), Wako, Saitama 351-0198, Japan}

\date{\today}

\begin{abstract}
    The nonlinear effect of a driving force in periodically driven quantum many-body systems can be systematically investigated 
    by analyzing the effective Floquet Hamiltonian. In particular, under an appropriate definition of the effective Hamiltonian, 
    simple driving forces may result in non-local interactions. 
    Here we consider a driven $ XXZ $ model to show that four-site interactions emerge owing to the driving force, which can 
    produce state-selective localization, a phenomenon where some limited Ising-like product states become fixed points of dynamics. 
    We first derive the effective Hamiltonian of a driven $ XXZ $ model on an arbitrary lattice for general spin $S$. 
    We then analyze in detail the case of a one-dimensional chain with $S=1/2$ as a special case, and find a condition 
    imposed on the cluster of four consecutive sites as a necessary and sufficient condition for the state of the whole system 
    to be localized. We construct such a localized state based on this condition and demonstrate it by numerical simulation of 
    the original time-periodic model.
\end{abstract}

\maketitle


\section{Introduction}\label{Section-1}


Periodically driven quantum many-body systems often have a structure that we can solve effectively as a time-independent system. This provides a hint for 
designing Hamiltonians with new quantum states as stationary states, which would not be realized in bare time-independent systems.
This technique is called \textit{Floquet engineering} in reference to the Floquet theory \cite{shirleySolutionSchrodingerEquation1965}, 
which is a general theory for time-periodic systems and has contributed to generation of new gauge fields in the optical lattice \cite{aidelsburgerExperimentalRealizationStrong2011,haukeNonAbelianGaugeFields2012,struckTunableGaugePotential2012,aidelsburgerRealizationHofstadterHamiltonian2013,miyakeRealizingHarperHamiltonian2013,struckEngineeringIsingXYSpinmodels2013,atalaObservationChiralCurrents2014,aidelsburgerArtificialGaugeFields2018} and deformation of the band structure of graphene \cite{guFloquetSpectrumTransport2011,kunduEffectiveTheoryFloquet2014,perez-piskunowFloquetChiralEdge2014,zhaiPhotoinducedTopologicalPhase2014,dehghaniOutofequilibriumElectronsHall2015,sentefTheoryFloquetBand2015,toppTopologicalFloquetEngineering2019,mciverLightinducedAnomalousHall2020}, for example.

The starting point of the Floquet engineering is to represent the periodically driven system in terms of a time-independent effective 
Hamiltonian. Using the Floquet theory, given a time-periodic Hamiltonian $ \hat{\mathcal{H}}(t) = \hat{\mathcal{H}}(t + T) $, we can 
construct a generator of time evolution over one cycle of $ \hat{\mathcal{H}}(t) $, which is called the \textit{Floquet Hamiltonian}. 
In the short-period limit $ T \to 0 $, or equivalently in the high-frequency limit $ \Omega \coloneqq 2\pi / T \to \infty $, the Floquet Hamiltonian coincides with the average Hamiltonian given by
\begin{align}
    \hat{\mathcal{H}}_{\mathrm{ave}} \coloneqq \frac{1}{T} \int_{0}^{T} dt\;\hat{\mathcal{H}}(t). \label{eq:average-Hamiltonian}
\end{align}
Therefore, the average Hamiltonian is often referred to as the effective (Floquet) Hamiltonian and is denoted by $ \hat{\mathcal{H}}_{\mathrm{eff}} $.

It is important, however, to note that we can in general obtain a number of effective Hamiltonians. In other words, for a given periodic 
Hamiltonian $ \hat{\mathcal{H}}(t) $, there are a generally infinite number of effective Hamiltonians
\begin{align}
    \hat{\mathcal{H}}_{\mathrm{eff}} = & \frac{1}{T} \int_{0}^{T}dt\;\hat{\mathcal{U}}(t) [\hat{\mathcal{H}}(t) - i\hbar \partial_{t}] \hat{\mathcal{U}}^{\dagger}(t), \label{eq:general-effective-Hamiltonian}
\end{align}
each of which is specified by the unitary transformation $ \hat{\mathcal{U}}(t) $. 
The average Hamiltonian in Eq.~\eqref{eq:average-Hamiltonian} is only a spacial case in which $ \hat{\mathcal{U}}(t) $ is the identity 
operator. Out of the infinitely large set $ \{ \hat{\mathcal{U}}(t) \} $, we should choose one that is appropriate to the purpose of research, 
especially when investigating nonlinear effects of driving forces in interacting many-body systems.

One example of such nonlinear effects is correlated tunneling in interacting bosons, which was predicted theoretically by Rapp \textit{et al.}\ \cite{rappUltracoldLatticeGases2012} for the Bose-Hubbard model with time-dependent on-site interactions and was demonstrated experimentally by Meinert \textit{et al.}\ \cite{meinertFloquetEngineeringCorrelated2016} using the corresponding cold-atom system. Correlated tunneling can only be described by $ \hat{\mathcal{H}}_{\mathrm{eff}} $ under a specific choice of $ \hat{\mathcal{U}}(t) $. The driven Bose-Hubbard model used in Refs.~\cite{rappUltracoldLatticeGases2012,meinertFloquetEngineeringCorrelated2016} is given by the time-dependent Hamiltonian
\begin{align}
    \hat{\mathcal{H}}(t) = -J \sum_{\pair{i}{j}} ( \hat{a}_{i}^{\dagger}\hat{a}_{j} + \mathrm{h.c.})  
    + \sum_{i} \frac{U(t)}{2} \hat{n}_{i}(\hat{n}_{i} - 1),\label{eq:driven-BHModel}
\end{align}
where $ \hat{a}_{i}^{\dagger} $ and $ \hat{a}_{i} $ are the bosonic creation and annihilation operators at site $i$, 
$ \hat{n}_{i} = \hat{a}_{i}^{\dagger} \hat{a}_{i} $ is the number operator, and the on-site interaction $ U(t) $ oscillates in time as $ U(t) = \bar{U} + \delta U \sin\Omega t $. Adopting the unitary transformation
\begin{align}
    \hat{\mathcal{U}}(t) = \exp\left[\frac{1}{2i}\frac{\delta U}{\hbar \Omega} \cos \Omega t \textstyle \sum_{i}\hat{n}_{i}(\hat{n}_{i} - 1)\right], \label{eq:unitary-for-BHModel}
\end{align}
we obtain the effective Hamiltonian
\begin{align}
    \hat{\mathcal{H}}_{\mathrm{eff}} = & -J \sum_{\pair{i}{j}} 
    \left( \hat{a}_{i}^{\dagger}
    \mathcal{J}_{0}(\hat{A}_{i, j})
    \hat{a}_{j} + \mathrm{h.c.} \right)
    + \sum_{i} \frac{\bar{U}}{2} \hat{n}_{i}(\hat{n}_{i} - 1),  \label{eq:effective-Bose-Hubbard-model}
\end{align}
where $ \mathcal{J}_{0}(x) $ is the zeroth-order Bessel function of the first kind and
\begin{align}
    \hat{A}_{i, j} \coloneqq (\delta U / \hbar \Omega)\times (\hat{n}_{i} - \hat{n}_{j})
\end{align}
is the scaled particle-number difference between sites $i$ and $j$.

In this Hamiltonian, the elementary processes describing the movement of particles are governed by effective bond operator
\begin{align}
    \hat{b}_{i, j}^{(\mathrm{eff})}\coloneqq \hat{a}_{i}^{\dagger} \mathcal{J}_{0}(\hat{A}_{i, j}) \hat{a}_{j} + \mathrm{h.c.}
\end{align}
for each bond $(i, j)$. Let $ n_{i} \coloneqq \bra{\Psi} \hat{n}_{i} \ket*{\Psi} $ be the expectation value of the number operator 
$ \hat{n}_{i} $ on site $ i $ under the state $ \ket*{\Psi} $. We can expect the dynamics through $ \hat{A}_{i, j} $ due to the bond operator 
$ \hat{b}_{i, j}^{(\mathrm{eff})}$ depending on the spatial distribution of $ n_{i} $. For example, for a bond $ (i, j) $, 
the process of changing the particle number distribution from $ (n_{i}, n_{j}) = (0, 1) $ to $ (n_{i}, n_{j}) = (1, 0) $ always takes place 
with a constant amplitude $ - J $ because $ \mathcal{J}_{0}(0) = 1 $. On the other hand, for the same bond, the process of changing 
from $ (n_{i}, n_{j}) = (1, 1) $ to $ (n_{i}, n_{j}) = (2, 0) $ takes place with an amplitude 
$ - J\times\mathcal{J}_{0}(\delta U / \hbar \Omega) $. 
When the value of $ (\delta U / \hbar \Omega) $ is set to a zero of the Bessel function $ \mathcal{J}_{0}(x) $, the latter process is strongly 
suppressed while the former process is intact. This complex state dependence is one of the features of correlated tunneling. 
Note that if we averaged the original Hamiltonian $ \hat{\mathcal{H}}(t) $ as in Eq.~\eqref{eq:average-Hamiltonian} 
by choosing the identity for $ \hat{\mathcal{U}}(t) $, we would only find a trivial time-independent part of $ \hat{\mathcal{H}}(t) $:
\begin{align}
    \hat{\mathcal{H}}_{\mathrm{ave}} = -J \sum_{\pair{i}{j}}\hat{a}_{i}^{\dagger}\hat{a}_{j}
    + \sum_{i} \frac{\bar{U}}{2} \hat{n}_{i}(\hat{n}_{i} - 1), \label{eq:effective-Bose-Hubbard-model-false}
\end{align}
from which we would not be able to derive correlated tunneling. 
This demonstrates that the choice of $ \hat{\mathcal{U}}(t) $ is essential for the observation of target phenomena.

When the total Hamiltonian is written as $ \hat{\mathcal{H}}(t) = \hat{\mathcal{H}}_{0} + \hat{\mathcal{V}}(t) $ such that $ \int_{0}^{2\pi/\Omega} dt\;\hat{\mathcal{V}}(t) = 0 $, we should utilize the unitary transformation in the form
\begin{align}
    \hat{\mathcal{U}}(t) = \mathcal{T}\exp[-\int_{}^{t}\frac{dt'}{i\hbar}\hat{\mathcal{V}}(t')], \label{eq:general-unitary-transformations}
\end{align}
which reduces to Eq.~\eqref{eq:unitary-for-BHModel} in the case of the Hamiltonian in Eq.~\eqref{eq:driven-BHModel}. 
Combining this with Eq.~\eqref{eq:general-effective-Hamiltonian}, we obtain the interaction term as 
\begin{align}
    \hat{\mathcal{V}}_{\mathrm{eff}} \coloneqq \frac{1}{T} \int_{0}^{T} dt\; [ \hat{\mathcal{U}}(t)\hat{\mathcal{H}}_{0}\hat{\mathcal{U}}^{\dagger}(t) - \hat{\mathcal{H}}_{0} ]. \label{eq:mapping-from-dynamical-to-static}
\end{align}
Note that the driving term in Eq.~\eqref{eq:driven-BHModel} acts on each site $ i $, whereas the resulting interaction in Eq.~\eqref{eq:effective-Bose-Hubbard-model} acts on each bond $ (i, j) $. This motivates us to investigate a possibility of finding, out of a driving two-body interaction, an effective long-range interaction of the form in Eq.~\eqref{eq:mapping-from-dynamical-to-static}.

Here in this study, we indeed find a four-site interaction out of an $ XXZ $ model in which the longitudinal exchange interaction is 
periodically driven with constant amplitude. In the resulting model, we observe a \textit{state-selective localization}, i.e., 
a limited number of Ising-like product states becoming fixed points of the dynamics generated by the effective Hamiltonian. 
This means that under the basis of the Ising-like product states, dynamics depends strongly on the initial state. 
Such an initial-state-dependent dynamics resembles the quantum scar recently studied extensively, 
which is one of the mechanisms preventing 
the thermal equilibration of quantum many-body 
systems~\cite{turnerQuantumScarredEigenstates2018,turnerWeakErgodicityBreaking2018}.

The rest of this paper is organized as follows. In Sec.~\ref{Section-2}, we derive the effective Hamiltonian of the driven $ XXZ $ model 
and show a general idea of state-selective localization through emergent long-range interactions. In Sec.~\ref{Section-3}, 
we analyze the localization condition in detail in the case of spin-$ 1/2 $ one-dimensional chain, providing a numerical demonstration. 
Finally, in Sec.~\ref{Section-4}, we conclude the paper with summary and discussion. 
The detailed derivation of the effective Hamiltonian is given in Appendix~\ref{Appendix-A}. 


\section{Emergent long-range interactions in a driven $ XXZ $ model}\label{Section-2}

We consider an $ XXZ $ model with periodically driven longuitudinal exchange interactions on an arbitrary lattice, whose Hamiltonian is given by
\begin{align}
    \hat{\mathcal{H}}(t) = - \frac{ J_{\perp} }{2} \sum_{\pair{i}{j}} (
        \hat{S}_{i}^{+} \hat{S}_{j}^{-}
        + \hat{S}_{i}^{-} \hat{S}_{j}^{+}
    ) - J_{\parallel}(t) \sum_{\pair{i}{j}} \hat{S}_{i}^{z} \hat{S}_{j}^{z},
    \label{eq:the-Original-Hamiltonian}
\end{align}
where $ J_{\perp} $ and $ J_{\parallel}(t) \equiv \bar{J}_{\parallel} + \delta J \cos\Omega t $ are the transverse and longitudinal 
components of the exchange interaction, respectively, and $ \{\hat{S}_{i}^{z,\pm}\} $ are the spin operators at site $i$, 
satisfying $ [\hat{S}_{i}^{+}, \hat{S}_{j}^{-}] = \delta_{i, j} \hbar \hat{S}_{i}^{z} $ and 
$ [\hat{S}_{i}^{z}, \hat{S}^{\pm}_{j}] = \pm \delta_{i, j} \hbar \hat{S}^{\pm}_{i} $. 
Here, the symbol $ \sum_{\pair{i}{j}} $ indicates summation over all the bonds on the lattice.

In the present case, for the unitary transformation in Eq.~\eqref{eq:general-unitary-transformations} we choose
\begin{align}
    \hat{\mathcal{U}}(t) = \exp\qty[\textstyle
    - i A \sin \Omega t \sum_{\pair{i}{j}} \hat{S}_{i}^{z} \hat{S}_{j}^{z}
    ],
\end{align}
where
\begin{align}
    A \coloneqq \frac{\delta J}{\hbar \Omega}
    \label{eq:dimensionless-amplitude}
\end{align}
is the dimensionless amplitude of the driving force and 
thus the effective Hamiltonian in Eq.~\eqref{eq:general-effective-Hamiltonian} takes the form
\begin{align}
    \hat{\mathcal{H}}_{\mathrm{eff}}(A) = - \frac{J_{\perp}}{2} \sum_{\pair{i}{j}} \left[
    \hat{S}_{i}^{+} \mathcal{J}_{0}(
      A \hat{Z}_{i, j}
    ) \hat{S}_{j}^{-}
    + (+ \leftrightarrow -)\right] - \bar{J}_{\parallel} \sum_{\pair{i}{j}} \hat{S}_{i}^{z} \hat{S}_{j}^{z}, \label{eq:effective-XXZ-model}
\end{align}
as described in Appendix~\ref{Appendix-A} for the derivation. Here,
\begin{align}
    \hat{Z}_{i, j} = \sum_{\langle k, \underline{i} \rangle} \hat{S}_{k}^{z} - \sum_{\langle k, \underline{j} \rangle} \hat{S}_{k}^{z} \label{eq:staggered-magnetization-operator}
\end{align}
is the local staggered magnetization operator around the bond $ (i, j) $. 
Note that $\sum_{\pair{k}{\underline{i}}} \hat{\mathcal{O}}_{k}$ denotes summation of $\hat{\mathcal{O}}_{k}$ 
over all sites $k$ that are connected to site $i$.

\begin{figure}
    \centering
    \includegraphics[width=0.90\linewidth]{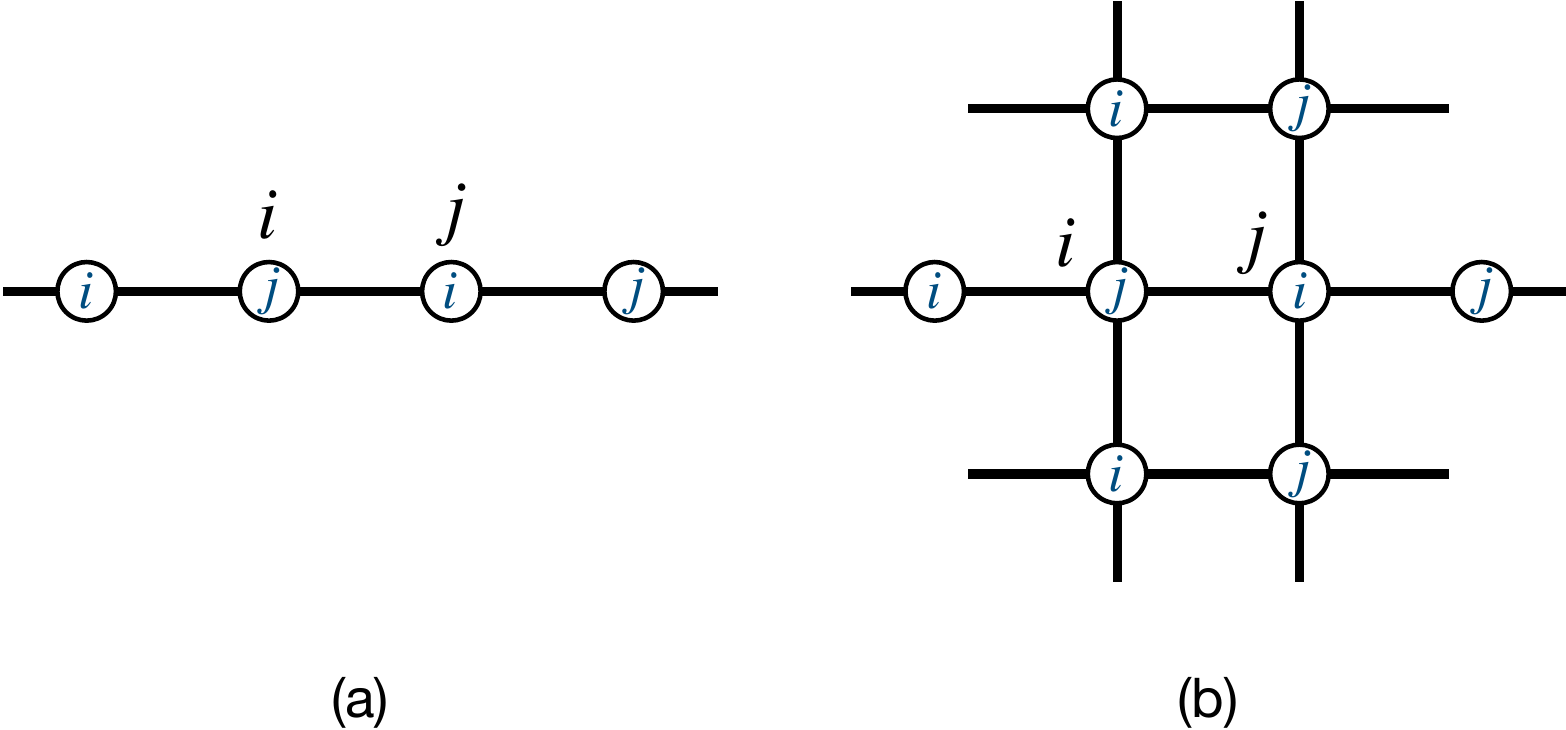}
    \caption{Range of action of $ \hat{Z}_{i, j} $ for a bond $ (i, j) $ in (a) a one-dimensional chain and (b) a two-dimensional square lattice. The gray letters inside the circles indicate the sites that are connected to either site $i$ or site $j$.}
    \label{fig:Z-illust}
\end{figure}

The operator $ \hat{Z}_{i, j} $ produces a new type of long-range interaction. First, the operator $ \hat{Z}_{i, j} $ involves many spins 
that depend on the underlying lattice structure. In fact, if the lattice is given on a hypercubic lattice of dimension $ d $, $ \hat{Z}_{i, j} $ is written as the sum of $ \hat{S}_{k}^{z} $ over $ 4d $ pieces of sites, since the coordination number is given by $ 2d $ (see Fig.~\ref{fig:Z-illust}). Second, $ \hat{Z}_{i, j} $ has the following property. 
Let the value of the parameter $ A $ be one of the innumerable zeros of the Bessel function $ \mathcal{J}_{0}(x) $, i.e., $ A_{\lambda} $ 
with $ \mathcal{J}_{0}(A_{\lambda}) = 0 $. If the operator $ \hat{Z}_{i, j} $ produces the eigenvalue $ \pm 1 $ in the argument of 
$ \mathcal{J}_{0}(A_{\lambda} \hat{Z}_{i, j}) $ when $ \hat{\mathcal{H}}_{\mathrm{eff}}(A_{\lambda}) $ acts on a quantum state 
$ \ket*{\Psi} $, then it holds that
\begin{align}
    \hat{\mathcal{H}}_{\mathrm{eff}}(A_{\lambda}) \ket*{\Psi} 
    = - \bar{J}_{\parallel} \sum_{\pair{i}{j}} \hat{S}_{i}^{z} \hat{S}_{j}^{z}\ket*{\Psi}. \label{eq:Ising-condition-0}
\end{align}
Since $ \hat{Z}_{i, j} $ is written in terms of $ \hat{S}_{k}^{z} $, let us choose the set of Ising-like product states as the representation 
basis hereafter. We can then classify all basis states into two types depending on the eigenvalues of $ \hat{Z}_{i, j} $, 
states for which the eigenvalue is either $ 1 $ or $ -1 $ and hence Eq.~\eqref{eq:Ising-condition-0} holds, 
and the other states for which Eq.~\eqref{eq:Ising-condition-0} does not hold.

We formulate it more specifically as follows. First, we break down the Hamiltonian into two terms:
\begin{align}
    \hat{\mathcal{H}}_{\mathrm{eff}}(A) = & \hat{\mathcal{H}}_{\mathrm{eff}}^{XY}(A) + \hat{\mathcal{H}}^{\mathrm{Ising}},
\end{align}
where
\begin{equation}
    \hat{\mathcal{H}}_{\mathrm{eff}}^{XY}(A) \coloneqq 
    - \frac{J_{\perp}}{2} \sum_{\pair{i}{j}} 
    \qty[
        \hat{S}_{i}^{+} \mathcal{J}_{0}(A \hat{Z}_{i, j}) \hat{S}_{j}^{-}
        + (+ \leftrightarrow -)
    ]\label{eq:effective-XY-hamiltonian}
\end{equation}
and
\begin{equation}
    \hat{\mathcal{H}}^{\mathrm{Ising}} \coloneqq 
    - \bar{J}_{\parallel} \sum_{\pair{i}{j}} \hat{S}_{i}^{z} \hat{S}_{j}^{z}. \label{eq:effective-Ising-hamiltonian}
\end{equation}
Let $ \mathfrak{H}^{\mathrm{Ising}} $ be the set of Ising-like product states that can be realized on a given lattice 
[see Eq.~\eqref{eq:Ising-like-product-states} for an example of $ S = 1/2 $]. Then an arbitrary state 
$ \ket*{\Psi} \in \mathfrak{H}^{\mathrm{Ising}} $ can be classified into two types depending on whether or not it satisfies the vanishing 
condition
\begin{align}
    \hat{\mathcal{H}}_{\mathrm{eff}}^{XY}(A_{\lambda})\ket*{\Psi} = 0. \label{eq:Ising-condition1}
\end{align}
Let $ \mathfrak{H}_{0}^{\mathrm{Ising}} $ be the set of the former, and $ \mathfrak{H}_{1}^{\mathrm{Ising}} $ be the set of the latter. 
The condition in Eq.~\eqref{eq:Ising-condition1} gives the decomposition into the direct sum 
$ \mathfrak{H}^{\mathrm{Ising}} = \mathfrak{H}_{0}^{\mathrm{Ising}} \oplus \mathfrak{H}_{1}^{\mathrm{Ising}} $. 
By casting the condition in Eq.~\eqref{eq:Ising-condition1} into the form
\begin{align}
    \exp\qty[
        \frac{1}{i\hbar}\hat{\mathcal{H}}_{\mathrm{eff}}(A_{\lambda})\times t
    ] \ket*{\Psi} \propto \ket*{\Psi}
\end{align}
for an arbitrary $ t\in\mathbb{R} $, we realize that the state $ \ket*{\Psi} \in \mathfrak{H}_{0}^{\mathrm{Ising}} $ is a fixed point of 
the dynamics generated by $ \hat{\mathcal{H}}_{\mathrm{eff}}(A_{\lambda}) $. In other words, $ \hat{\mathcal{H}}_{\mathrm{eff}}(A) $ 
generates dynamics such that only the states $ \ket*{\Psi} \in \mathfrak{H}_{0}^{\mathrm{Ising}} $ are selectively localized.

\section{Spin-$ 1/2 $ one-dimensional chain}\label{Section-3}

In order to investigate the state-selective localization more specifically, we now analyze an $ S = 1/2 $ chain of length $ L $ under the 
periodic boundary conditions. The two terms in Eqs.~\eqref{eq:effective-XY-hamiltonian} and \eqref{eq:effective-Ising-hamiltonian} 
constituting the effective Hamiltonian $\hat{\mathcal{H}}_{\mathrm{eff}}(A) $ in Eq.~\eqref{eq:effective-XXZ-model} now reads
\begin{equation}
    \hat{\mathcal{H}}_{\mathrm{eff}}^{XY}(A)  = 
    - \frac{J_{\perp}}{2} \sum_{i = 1}^{L} 
    \qty[
        \hat{S}_{i}^{+} \mathcal{J}_{0}(A \hat{Z}_{i, i + 1}) \hat{S}_{i + 1}^{-}
        + (+ \leftrightarrow -)
    ]\label{eq:the-effective-XY-hamiltonian1}
\end{equation}
and 
\begin{equation}
    \hat{\mathcal{H}}^{\mathrm{Ising}}  = 
    - \bar{J}_{\parallel} \sum_{i = 1}^{L} \hat{S}_{i}^{z} \hat{S}_{i + 1}^{z},
\end{equation}
respectively, where $ \hat{\vb*{S}}_{L + 1} = \hat{\vb*{S}}_{1} $ and
\begin{align}
    \hat{Z}_{i, i + 1} = \hat{S}_{i - 1}^{z} - \hat{S}_{i}^{z} + \hat{S}_{i + 1}^{z} - \hat{S}_{i + 2}^{z}.
\end{align}

Let us denote each element of $ \mathfrak{H}^{\mathrm{Ising}} $ as
\begin{align}
    \ket*{\Psi_{\vb*{m}}} = \bigotimes_{l = 1}^{L} \ket*{m_{i}} = \ket*{m_{1}, m_{2}, \ldots, m_{L}}, \label{eq:Ising-like-product-states}
\end{align}
where $ \ket*{m_{j}} $ is either of the local eigenstates $ \ket*{\uparrow}, \ket*{\downarrow} $ defined by
\begin{align}
    \hat{S}_{j}^{z} \ket*{\uparrow} & = \frac{1}{2} \ket*{\uparrow}, \quad \hat{S}_{j}^{z} \ket*{\downarrow} = - \frac{1}{2} \ket*{\downarrow}
\end{align}
and $ \vb*{m} = (m_{1}, m_{2}, \ldots, m_{L}) $ in the label of the state $ \ket*{\Psi_{\vb*{m}}} $. 
In this case, the condition in Eq.~\eqref{eq:Ising-condition1} for $ \ket*{\Psi} $ is equivalent to the condition
\begin{align}
    \hat{b}_{i, i + 1}(A_{\lambda})\ket*{\Psi_{\vb*{m}}} = 0 \label{eq:semi-local-Ising-condition}
\end{align}
for every bond $ (i, i + 1) $, where we introduced the effective bond operator
\begin{align}
    \hat{b}_{i, i + 1}(A) \coloneqq \hat{S}_{i}^{+}\mathcal{J}_{0}(A\hat{Z}_{i, i + 1})\hat{S}_{i + 1}^{-}
    + (+ \leftrightarrow -)
    \label{eq:bond-operator}
\end{align}
Since the bond operator $ \hat{b}_{i, i + 1} $ acts on the four spins at sites $ i - 1 $, $ i $, $ i + 1 $, and $ i + 2$, 
we can focus on the $ 16 $ pieces of four-site states
\begin{align}
    \ket*{\psi_{i, i + 1}} = \ket*{m_{i - 1}, m_{i}, m_{i + 1}, m_{i + 2}},\label{eq:four-site-state}
\end{align}
out of the whole state $ \ket*{\Psi_{\vb*{m}}} $ given by
\begin{align}
    \ket*{\Psi_{\vb*{m}}} = \ket*{m_{1} \ldots m_{i - 2}} \otimes 
    \ket*{\psi_{i, i + 1}} \otimes
    \ket*{m_{i + 3} \ldots m_{L}},\label{eq:partial-bond-state}
\end{align}
reducing the condition in Eq.~\eqref{eq:semi-local-Ising-condition} to
\begin{align}
    \hat{b}_{i, i + 1}(A_{\lambda})\ket*{\psi_{i, i + 1}} = 0. \label{eq:local-Ising-condition}
\end{align}

First, we can classify the 16 four-site states into two groups: a group of 8 states satisfying 
$ \ket*{m_{i + 1}} = \ket*{\uparrow} $ and 
a group of 8 states satisfying $ \ket*{m_{i + 1}} = \ket*{\downarrow} $. 
For any state $ \ket{\psi_{i, i + 1}} $ in the former group, it holds that
\begin{align}
    \hat{b}_{i, i + 1}(A) \ket{\psi_{i, i + 1}}
    = \hat{S}_{i}^{+}
    \mathcal{J}_{0}(A\hat{Z}_{i, i + 1})
    \hat{S}_{i + 1}^{-} \ket{\psi_{i, i + 1}}.
\end{align}
In this case, since $ \hat{S}_{i + 1}^{-} \ket{\psi_{i, i + 1}} $ is also one of the 16 four-site states, it is an eigenstate of $ \mathcal{J}_{0}(A\hat{Z}_{i, i + 1}) $, and thus we have
\begin{align}
    \hat{S}_{i}^{+}
    \mathcal{J}_{0}(A\hat{Z}_{i, i + 1})
    \hat{S}_{i + 1}^{-} \ket{\psi_{i, i + 1}}
    \propto \hat{S}_{i}^{+}
    \hat{S}_{i + 1}^{-} \ket{\psi_{i, i + 1}}
\end{align}
with the corresponding eigenvalue as the proportionality factor. In the same way, for any state $ \ket{\psi_{i, i + 1}} $ in the latter group, since it holds that
\begin{align}
    \hat{b}_{i, i + 1}(A) \ket{\psi_{i, i + 1}}
    = \hat{S}_{i}^{-}
    \mathcal{J}_{0}(A\hat{Z}_{i, i + 1})
    \hat{S}_{i + 1}^{+} \ket{\psi_{i, i + 1}}.
\end{align}
and since $ \hat{S}_{i + 1}^{+} \ket{\psi_{i, i + 1}} $ is an eigenstate of $ \mathcal{J}_{0}(A\hat{Z}_{i, i + 1}) $, we have
\begin{align}
    \hat{S}_{i}^{-}
    \mathcal{J}_{0}(A\hat{Z}_{i, i + 1})
    \hat{S}_{i + 1}^{+} \ket{\psi_{i, i + 1}}
    \propto \hat{S}_{i}^{-}
    \hat{S}_{i + 1}^{+} \ket{\psi_{i, i + 1}}.
\end{align}
Putting these two together, we obtain
\begin{align}
    \hat{b}_{i, i + 1}(A) \ket{\psi_{i, i + 1}}
    = \bar{\mathcal{J}}
    \times \hat{b}_{i, i + 1}(0) \ket{\psi_{i, i + 1}} \label{eq:misterious-equation}
\end{align}
with the proportionality coefficient $ \bar{\mathcal{J}} $.

\begin{figure}
    \centering
   \includegraphics[width=0.9\linewidth]{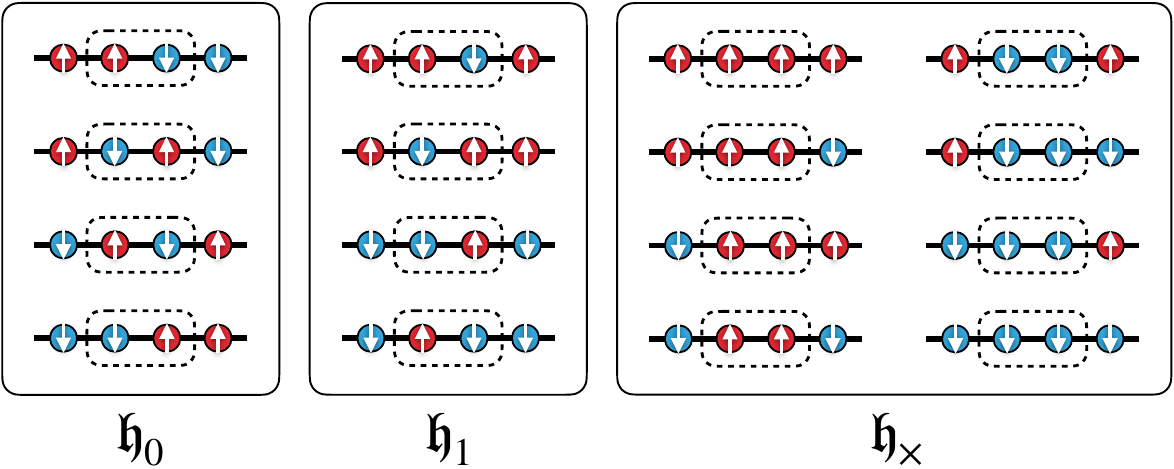}
    \caption{Classification of $2^{4} = 16$ possible Ising-like product states in four consecutive sites (a cluster). The three groups $ \mathfrak{h}_{0} $, $ \mathfrak{h}_{1} $, and $ \mathfrak{h}_{\times} $ shown in the figure correspond to the states given by Eqs.~\eqref{eq:frak-small-h0}, \eqref{eq:frak-small-h1}, and \eqref{eq:frak-small-hx}, respectively. In each cluster state, the two sites enclosed in a dashed rectangle correspond to the bond $ (i, i + 1) $.}
    \label{fig:16-states}
\end{figure}

We then classify the 16 four-site states further into three groups (also see Fig.~\ref{fig:16-states}): 
\begin{align}
    \mathfrak{h}_{0} \coloneqq & \{
        \ket*{\uparrow\uparrow\downarrow\downarrow}, 
        \ket*{\uparrow\downarrow\uparrow\downarrow}, 
        \ket*{\downarrow\uparrow\downarrow\uparrow}, 
        \ket*{\downarrow\downarrow\uparrow\uparrow}
    \}, \label{eq:frak-small-h0}\\
    \mathfrak{h}_{1} \coloneqq & \{
        \ket*{\uparrow\uparrow\downarrow\uparrow}, 
        \ket*{\uparrow\downarrow\uparrow\uparrow}, 
        \ket*{\downarrow\downarrow\uparrow\downarrow}, 
        \ket*{\downarrow\uparrow\downarrow\downarrow}
    \}, \label{eq:frak-small-h1}\\
    \mathfrak{h}_{\times} \coloneqq & \{
        \ket*{m_{i - 1}\uparrow\uparrow m_{i + 2}}, 
        \ket*{m_{i - 1}\downarrow\downarrow m_{i + 2}} 
    \}_{m_{i - 1}, m_{i + 2} = \uparrow, \downarrow}.
    \label{eq:frak-small-hx}
\end{align}
The behavior of $ \bar{\mathcal{J}} $, $ \hat{b}_{i, i + 1}(0) \ket{\psi_{i, i + 1}} $, and $ \bar{\mathcal{J}}
\times \hat{b}_{i, i + 1}(0) \ket{\psi_{i, i + 1}} $ on the right-hand side of Eq.~\eqref{eq:misterious-equation} is shown in Table~\ref{table:values}. 
Therefore, setting $ A = A_{\lambda} $ yields
\begin{widetext}
\begin{align}
    \hat{b}_{i, i + 1}(A_{\lambda}) \ket*{\psi_{i, i + 1}} = \begin{cases}
        0 & \qq{$ \ket*{\psi_{i, i + 1}} \in  \mathfrak{h}_{0} \oplus \mathfrak{h}_{\times}$}\\
        \hat{b}_{i, i + 1}(0) \ket*{\psi_{i, i + 1}} & \qq{$ \ket*{\psi_{i, i + 1}} \in \mathfrak{h}_{1} $}
    \end{cases}.\label{eq:local-Ising-condition-final}
\end{align}
\end{widetext}
From the above, we find that the four-site state $ \ket*{\psi_{i, i + 1}} $ satisfying the local vanishing condition 
in Eq.~\eqref{eq:local-Ising-condition} is given by the union set $ \mathfrak{h}_{0} \oplus \mathfrak{h}_{\times} $. 
On this basis, we can also construct a state $ \ket*{\Psi_{\vb*{m}}} $ that satisfies the global vanishing condition 
in Eq.~\eqref{eq:Ising-condition1}. Next, we shall numerically verify that a state $ \ket*{\Psi_{\vb*{m}}} $ satisfying 
Eq.~\eqref{eq:Ising-condition1} is indeed robust against the corresponding periodic drive. 

\begin{table}
    \centering
    \caption{The right-hand side of Eq.~\eqref{eq:misterious-equation} for three cases of 
    $ \ket{\psi_{i, i + 1}} \in \mathfrak{h}_{0}, \mathfrak{h}_{1}, \mathfrak{h}_{\times} $.}
    \begin{tabular}{l|lll}\hline \hline
    \; & $ \bar{\mathcal{J}} $ & $ \hat{b}_{i, i + 1}(0) \ket{\psi_{i, i + 1}} $ & $ \bar{\mathcal{J}} \times \hat{b}_{i, i + 1}(0) \ket{\psi_{i, i + 1}} $ \\
    \hline
    $ \ket{\psi_{i, i + 1}}\in \mathfrak{h}_{0} $ & $ \mathcal{J}_{0}(A) $ & $ \in\mathfrak{h}_{0} $ & $ \in \mathcal{J}_{0}(A) \mathfrak{h}_{0} $\\
    $ \ket{\psi_{i, i + 1}}\in \mathfrak{h}_{1} $ & $ 1 $ & $ \in\mathfrak{h}_{1} $ & $ \in \mathfrak{h}_{1} $\\
    $ \ket{\psi_{i, i + 1}}\in \mathfrak{h}_{\times} $ & It depends. & $ 0 $ & $ 0 $ \\
    \hline
    \end{tabular}
    \label{table:values}
\end{table}

    Let us consider the following two product states:
\begin{align}
    \ket*{\mathrm{A}_{0}} = & \ket*{
        \downarrow\downarrow\downarrow\downarrow\downarrow\downarrow\downarrow\underline{
            \downarrow\uparrow
        }\uparrow\uparrow\uparrow\uparrow\uparrow\uparrow\uparrow
    }, \label{eq:A0}\\
    \ket*{\mathrm{A}_{1}} = & \ket*{
        \downarrow\downarrow\downarrow\downarrow\downarrow\downarrow\downarrow\underline{
            \uparrow\downarrow
        }\uparrow\uparrow\uparrow\uparrow\uparrow\uparrow\uparrow
    } \label{eq:A1}
\end{align}
on an $ L = 16 $ finite-size chain with the periodic boundary conditions. 
The former is a state with one domainwall at the bond $ (8, 9) $ and is contained in $ \mathfrak{H}_{0}^{\mathrm{Ising}} $ 
because $ \ket*{\psi_{n, n + 1}} \in \mathfrak{h}_{\times} \oplus \mathfrak{h}_{0} $ holds for every bond 
$ (n, n + 1) \in E $. 
The latter is a state in which the spins at the bond $ (8, 9) $, the position indicated by the underlines in 
Eqs.~\eqref{eq:A0} and \eqref{eq:A1}, are flipped and is contained in $ \mathfrak{H}_{1}^{\mathrm{Ising}} $ because the bonds 
$ (7, 8) $ and $ (9, 10) $ satisfy $ \ket*{\psi_{7, 8}} \in \mathfrak{h}_{1} $ and $ \ket*{\psi_{9, 10}} \in \mathfrak{h}_{1} $.

These two states $ \ket*{\mathrm{A}_{0}} $ and $ \ket*{\mathrm{A}_{1}} $ only differ in the partial state at the bond $ (8, 9) $, which may be negligible in the thermodynamic limit $ L \to \infty $. The slight difference, nevertheless, generates 
largely different time-evolution dynamics due to the fact that one state belongs to $ \mathfrak{H}_{0}^{\mathrm{Ising}} $ and the other 
state belongs to $ \mathfrak{H}_{1}^{\mathrm{Ising}} $. 
To explore this quantitatively, we calculate the value of $ S_{n}^{z}(t) \coloneqq \bra{\Psi(t)} \hat{S}_{n}^{z} \ket{\Psi(t)} $ and 
the half-chain entanglement entropy per a lattice site given by
\begin{align}
    \sigma_{L / 2}(t)\coloneqq
    \frac{1}{L} \mathrm{Tr} \qty[
        - \hat{\rho}_{L / 2}(t)
        \log \hat{\rho}_{L / 2}(t)
    ],
\end{align}
where
\begin{align}
    \hat{\rho}_{L / 2}(t)  \coloneqq
    \mathrm{Tr}_{1\leq n \leq L  / 2} \qty[
        \ket{\Psi(t)}\bra{\Psi(t)}
    ].
\end{align}
The state vector at time $ t $ is given by
\begin{align}
    \ket*{\Psi(t)} = \mathcal{T}\exp\qty[
        \int_{0}^{t}\frac{dt'}{i\hbar}
        \hat{\mathcal{H}}(t')
    ] \ket*{\Psi(0)},
\end{align}
which is numerically calculated by the package 
QuSpin~\cite{weinbergQuSpinPythonPackage2017b,weinbergQuSpinPythonPackage2019b}. 
We set the parameters in the Hamiltonian $ \hat{\mathcal{H}}(t) $ for which the effective Hamiltonian 
$ \hat{\mathcal{H}}_{\mathrm{eff}}(A_{\lambda}) $ well approximates the dynamics and the condition 
$ \mathcal{J}_{0}(A_{\lambda}) \simeq 0 $ is satisfied. 

\begin{figure*}
    \centering
   \includegraphics[width=0.9\linewidth]{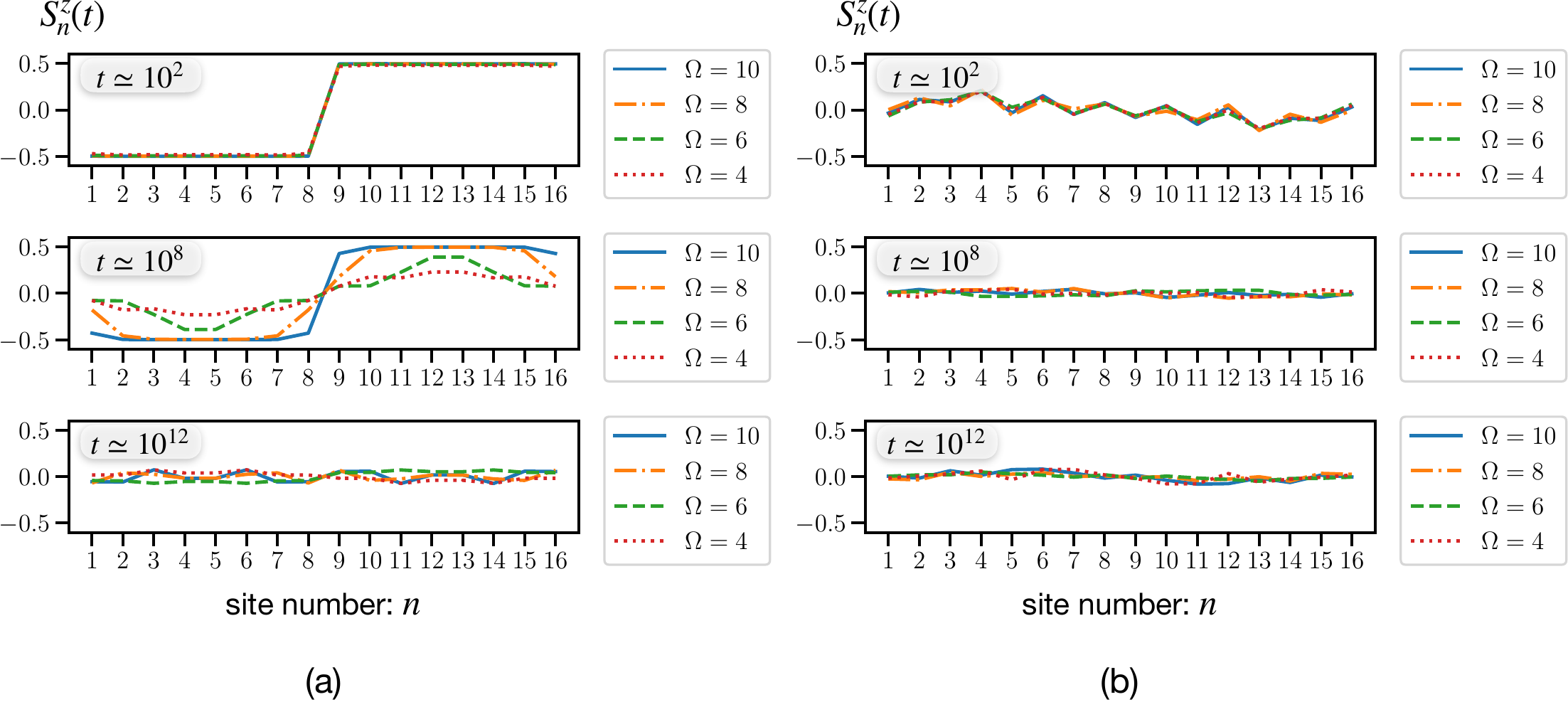}
    \caption{Numerical results of $ S_{n}^{z}(t) $ for three different times (in unit of $|J_{\parallel}|$) with the initial states (a) $ \ket*{\Psi(0)} = \ket*{\mathrm{A}_{0}} $ 
    and (b) $ \ket*{\Psi(0)} = \ket*{\mathrm{A}_{1}} $. 
    Here, the value of $ S_{n}^{z}(t)=0.5 $ corresponds to 
    the local spin state $ \ket*{\uparrow} $, while the value of $ S_{n}^{z}(t)=-0.5 $ corresponds to the local spin 
    state $ \ket*{\downarrow} $. 
    We set  $A = 2.4048$ and hence $ \mathcal{J}_{0}(A) \simeq 0 $ holds. The other parameters are $J_{\parallel} = -1 $ and 
    $ J_{\perp} = -0.75 $ with $ \hbar = 1$. 
    }
    \label{fig:typeA-Sz}
\end{figure*}

Figure~\ref{fig:typeA-Sz} shows the results of the spatial profile of $ S_{n}^{z}(t) $ at several different times for the initial states 
$ \ket*{\mathrm{A}_{0}} $ and $ \ket*{\mathrm{A}_{1}} $ and for the frequencies $ \Omega = 10, 8, 6, 4 $. 
These results demonstrate that 
the slight difference of the spin configuration in the initial states $ \ket*{\mathrm{A}_{0}} $ and $ \ket*{\mathrm{A}_{1}} $ 
makes a significant difference in the time evolution; the spin configuration remains robust and almost intact for $ \ket*{\mathrm{A}_{0}} $ at the beginning, but decays quickly for $ \ket*{\mathrm{A}_{1}} $. Although the initial spin configuration for $ \ket*{\mathrm{A}_{0}} $ is also eventually destroyed, the robustness of $ \ket*{\mathrm{A}_{0}} $ is a direct consequence of the fact that the initial state $ \ket*{\mathrm{A}_{0}} $ is at the very vicinity of a fixed point of the dynamics.

\begin{figure*}
    \centering
   \includegraphics[width=0.8\linewidth]{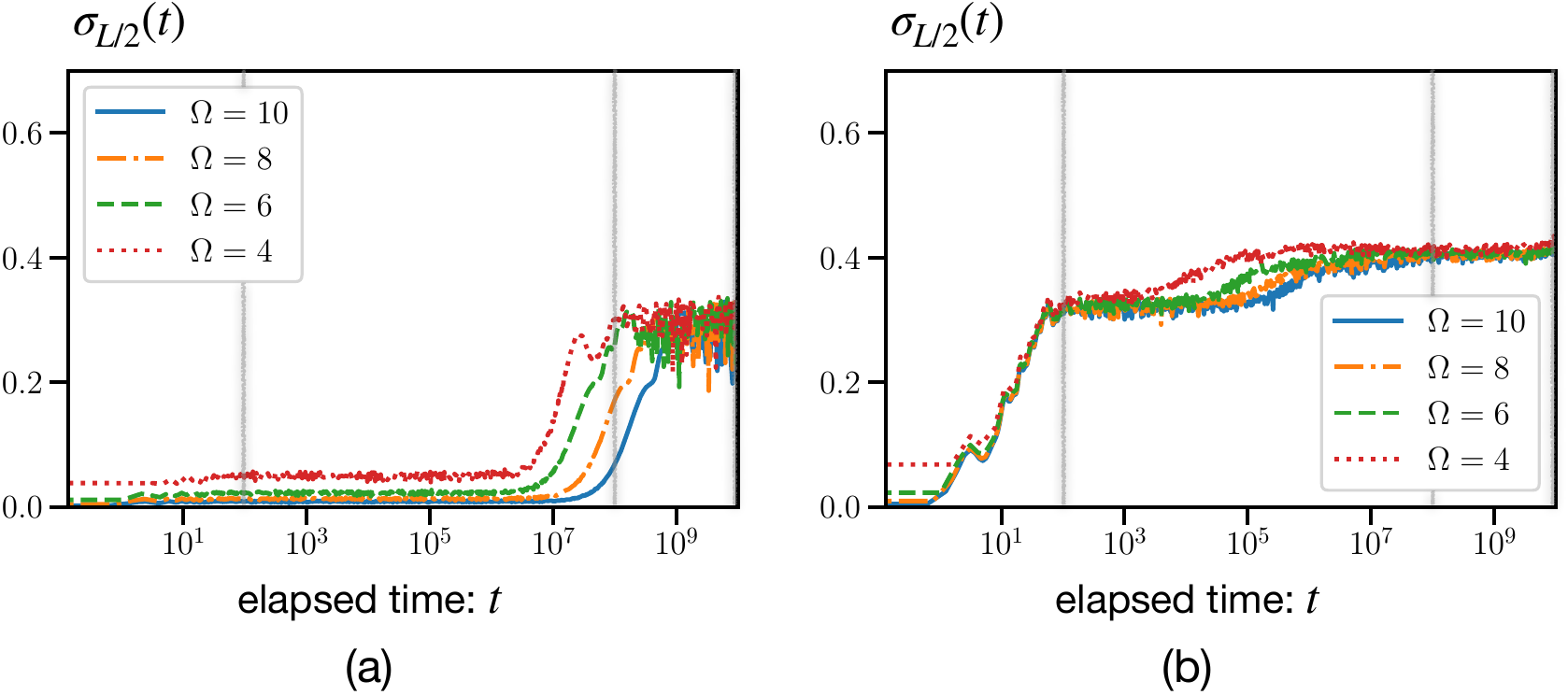}
    \caption{
        Numerical results of the time evolution of $ \sigma_{L / 2}(t) $ for the initial states (a) $ \ket*{\Psi(0)} = \ket*{\mathrm{A}_{0}} $ 
    and (b) $ \ket*{\Psi(0)} = \ket*{\mathrm{A}_{1}} $. 
    The parameters of the calculations are the same as in Fig.~\ref{fig:typeA-Sz} and the unit of time is $|J_{\parallel}|$. 
    The gray vertical lines indicate the three different times at which $ S_{n}^{z}(t) $ is evaluated in Fig.~\ref{fig:typeA-Sz}.
    }
    \label{fig:typeA-Ent}
\end{figure*}

Figure~\ref{fig:typeA-Ent} shows the time evolution of $ \sigma_{L / 2}(t) $ for the same set of the initial states and 
for the same frequencies $ \Omega $. Figure~\ref{fig:typeA-Ent}(a) suggests that when the initial state satisfies 
$ \hat{\mathcal{H}}_{\mathrm{eff}}^{XY}(A_{\lambda})\ket*{\Psi} \simeq 0 $, the initial state essentially remains intact until a certain 
time that depends on $ \Omega $ and can be delayed as late as possible with increasing $ \Omega $. 
Figure~\ref{fig:typeA-Ent}(b) suggests that when the initial state satisfies 
$ \hat{\mathcal{H}}_{\mathrm{eff}}^{XY}(A_{\lambda})\ket*{\Psi} \ne 0 $, 
$ \sigma_{L / 2}(t) $ increases rather rapidly at a certain time that does not 
depend on $ \Omega $. In addition, there is another relaxation process at a later time 
that depends on $ \Omega $ and can be delayed as late as possible with increasing $ \Omega $. 
This feature of $ \Omega $-dependent relaxation found in Figs.~\ref{fig:typeA-Ent}(a) and \ref{fig:typeA-Ent}(b)
is similar to the prethermalization and subsequent relaxation to the infinite temperature state reported in some Floquet systems.

However, as reported in Ref.~\cite{machadoExponentiallySlowHeating2019}, the value of entanglement entropy corresponding to 
the infinite temperature state usually takes the universal value $ \sigma_{L / 2} \sim \ln 2 - 1 / L $. 
This is different form the values of entanglement entropy estimated at the largest time available in Fig.~\ref{fig:typeA-Ent}, which 
are in fact both far from that value. Besides, there is no guarantee that these values of the entanglement entropy at the largest time 
in Fig.~\ref{fig:typeA-Ent} is actually the value after the full relaxation. Thus, one of two possible conclusions can be drawn from our 
results: either a new and unprecedented relaxation phenomenon occurs, or no relaxation is occurring at all. 
Verifying this would require additional computational resources and more extended simulations on larger systems. 
Therefore, we leave this issue in a future study. 

\begin{figure*}
    \centering
   \includegraphics[width=0.9\linewidth]{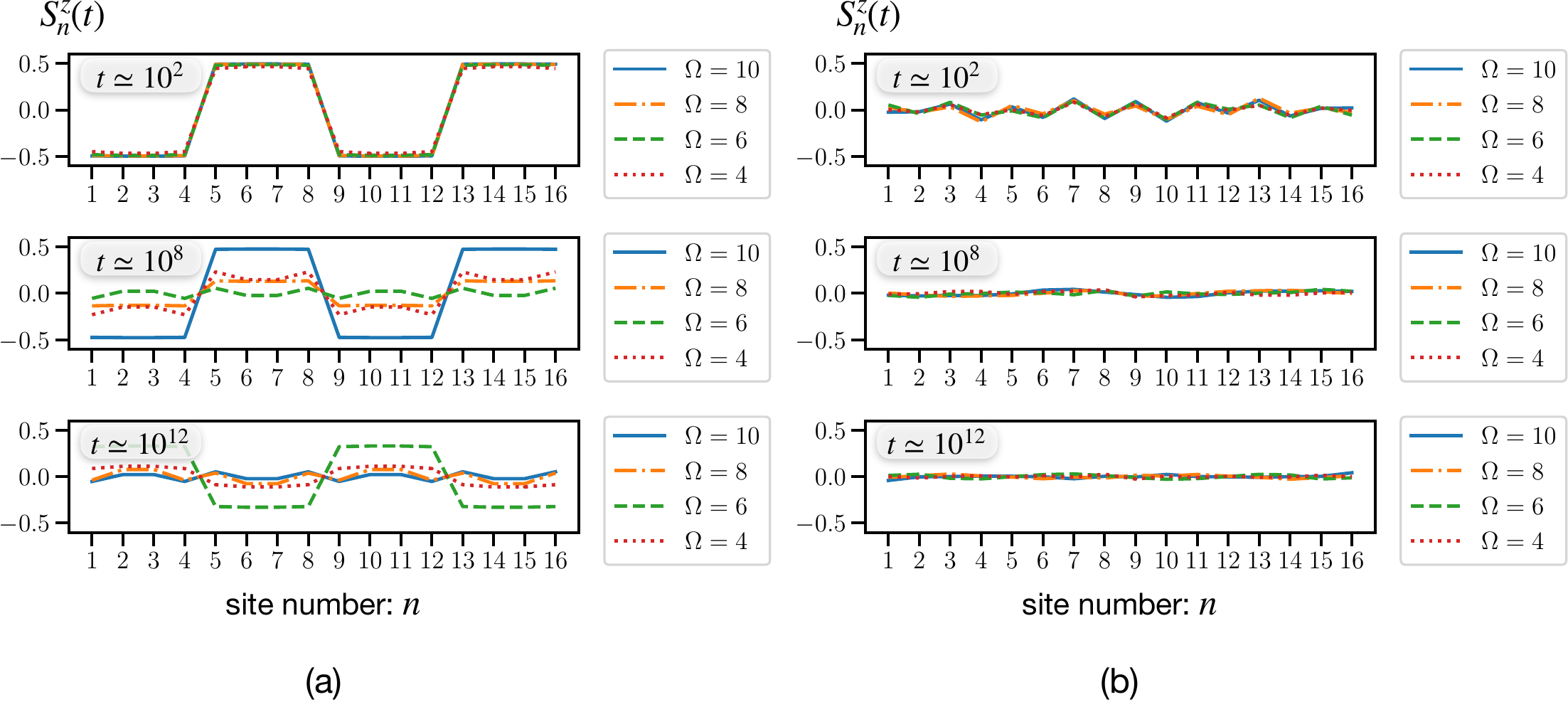}
    \caption{
    Same as Fig.~\ref{fig:typeA-Sz} but for the initial states (a) $ \ket*{\Psi(0)} = \ket*{\mathrm{B}_{0}} $ 
    and (b) $ \ket*{\Psi(0)} = \ket*{\mathrm{B}_{1}} $. 
    }
    \label{fig:typeB-Sz}
\end{figure*}

\begin{figure*}
    \centering
   \includegraphics[width=0.8\linewidth]{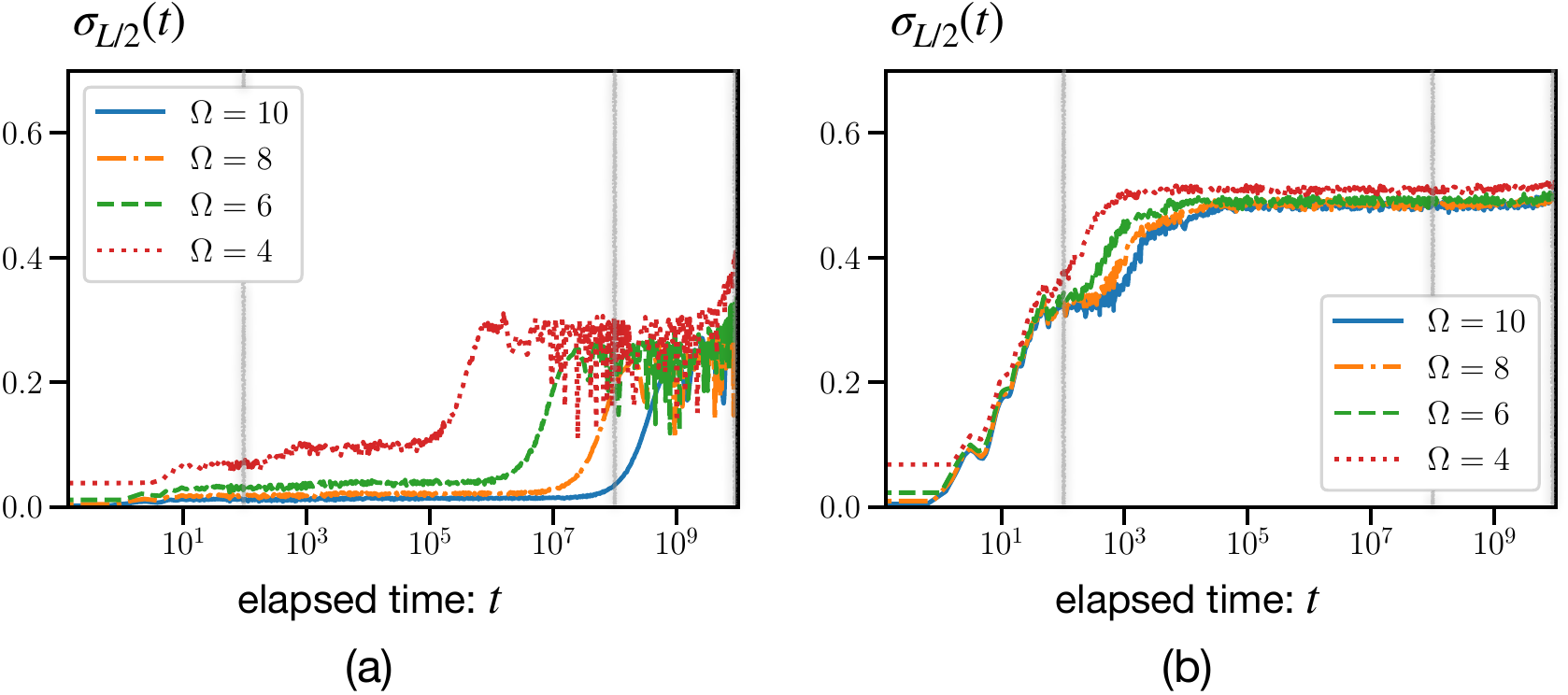}
    \caption{
    Same as Fig.~\ref{fig:typeA-Ent} but for the initial states (a) $ \ket*{\Psi(0)} = \ket*{\mathrm{B}_{0}} $ 
    and (b) $ \ket*{\Psi(0)} = \ket*{\mathrm{B}_{1}} $. 
    }
    \label{fig:typeB-Ent}
\end{figure*}

We also perform the same numerical calculations for the following two product states as the initial states: 
\begin{align}
    \ket*{\mathrm{B}_{0}} = & \ket*{
        \downarrow\downarrow\downarrow\downarrow
        \uparrow\uparrow\uparrow\underline{
            \uparrow\downarrow
        }\downarrow\downarrow\downarrow
        \uparrow\uparrow\uparrow\uparrow
    }, \label{eq:B0}\\
    \ket*{\mathrm{B}_{1}} = & \ket*{
        \downarrow\downarrow\downarrow\downarrow
        \uparrow\uparrow\uparrow\underline{
            \downarrow\uparrow
        }\downarrow\downarrow\downarrow
        \uparrow\uparrow\uparrow\uparrow
    }. \label{eq:B1}
\end{align}
These states also form a set of states where $\ket*{\mathrm{B}_{0}} $ belongs to $ \mathfrak{h}_{0} $ and 
$\ket*{\mathrm{B}_{1}} $ belongs $ \mathfrak{h}_{1} $ 
with a slight difference at the bond $ (8, 9) $, indicated by the underlines in Eqs.~\eqref{eq:B0} and \eqref{eq:B1}. 
Figure~\ref{fig:typeB-Sz} shows the numerical results for the spatial profile of $ S_{n}^{z}(t) $ at three different times 
for the initial states $ \ket*{\mathrm{B}_{0}} $ and $ \ket*{\mathrm{B}_{1}} $ and for the frequencies $ \Omega = 10, 8, 6, 4 $. 
Figure \ref{fig:typeB-Ent} shows the time evolution of $ \sigma_{L / 2}(t) $ for the same set of initial states and for 
the same frequencies $ \Omega $.
The behaviors shown in Figs.~\ref{fig:typeB-Sz} and \ref{fig:typeB-Ent} are qualitatively the same as those in 
Figs.~\ref{fig:typeA-Sz} and \ref{fig:typeA-Ent}, respectively. 

Our results imply that by simply flipping a pair of spins on a single bond we can switch the whole state back and forth 
between $ \mathfrak{H}_{0}^{\mathrm{Ising}} $ and $ \mathfrak{H}_{1}^{\mathrm{Ising}} $. 
Hence, it is also possible to infer from the measurement of $ S_{n}^{z}(t) $ whether the initial state $ \ket*{\Psi(0)} $ belongs 
to $ \mathfrak{H}_{0}^{\mathrm{Ising}} $ or $ \mathfrak{H}_{1}^{\mathrm{Ising}} $, provided that the value of $A$ is appropriately 
tuned.

\section{Conclusion and Discussion}\label{Section-4}

We have shown that driven longitudinal exchange interactions in the $ XXZ $ model lead to long-range transverse exchange 
interactions in the Floquet picture, and result in a state-selective localization in which limited Ising-like product states are fixed 
points of the dynamics. Especially in the one-dimensional case with spin $ S = 1/2 $ and arbitrary length $ L $, we have shown 
the long-range transverse interactions reduce to four-site interactions and specified the condition for a given Ising-like product 
state to be the fixed point. We have also shown some examples of such a localizable product state and numerically verified 
that these are actually fixed points of the dynamics by demonstrating quantitatively different dynamics for two different 
initial product states with only a slightly different spin configuration on a single bond. 

The driving protocol presented in this study may be experimentally realized in magnetic insulators by a technique of controlling 
magnetism with ultrafast electric fields~\cite{mentinkManipulatingMagnetismUltrafast2017}. 
In addition to such experimental approaches, it is also interesting to investigate the state-selective localization as a quantum 
scar~\cite{turnerQuantumScarredEigenstates2018,turnerWeakErgodicityBreaking2018}, which prevents thermal equilibration in 
isolated quantum many-body systems. The state-selective localization proposed in this study has a common feature to the quantum 
scar in that the dynamics is strongly dependent on the initial state. It is desirable that the model given in this study is 
properly extended to describe experimental situations and to investigate the quantum scar in more details.

\acknowledgements
We are grateful to Takashi Oka for valuable comments. K.S is financially supported by the Junior Research Associate (JRA) program in RIKEN. S.Y is financially supported in part by JSPS KAKENHI with Grant No.~JP18H01183.

\appendix

\section{Derivation of $ \hat{\mathcal{H}}_{\mathrm{eff}}(A) $ in Eq.~\eqref{eq:effective-XXZ-model}}\label{Appendix-A}

In this appendix, we present a detailed derivation of $ \hat{\mathcal{H}}_{\mathrm{eff}}(A) $ in Eq.~\eqref{eq:effective-XXZ-model}. 
First, we decompose the original time-dependent Hamiltonian $\hat{\mathcal{H}}(t)$ given in Eq.~\eqref{eq:the-Original-Hamiltonian} 
into $ \hat{\mathcal{H}}(t) = \hat{\mathcal{H}}_{0} + \hat{\mathcal{V}}(t) $ such that 
$ \int_{0}^{2\pi / \Omega} dt \;\hat{\mathcal{V}}(t) = 0 $ holds. Then, for
\begin{equation}
    \hat{\mathcal{H}}_{0} =  - \frac{J_{\perp} }{2} \sum_{\pair{i}{j}} ( \hat{S}_{i}^{+} \hat{S}_{j}^{-} + \hat{S}_{i}^{-} \hat{S}_{j}^{+}) - \bar{J_{\parallel}} \sum_{\pair{i}{j}} \hat{S}_{i}^{z} \hat{S}_{j}^{z} 
\end{equation}
and
\begin{equation}
    \hat{\mathcal{V}}(t) =  - \delta J \cos \Omega t \sum_{\pair{i}{j}} \hat{S}_{i}^{z} \hat{S}_{j}^{z},
\end{equation}
we find the unitary transformation in Eq.~\eqref{eq:general-unitary-transformations} in the form
\begin{align}
    \hat{\mathcal{U}}(t) = & \mathcal{T}\exp[-\int_{}^{t}\frac{dt'}{i\hbar}\hat{\mathcal{V}}(t')]\\
    = & \textstyle\exp\qty[
        -iA \sin \Omega t \sum_{\pair{k}{l}}\hat{S}_{k}^{z}\hat{S}_{l}^{z}
    ],
\end{align}
where $ A $ is given in Eq.~\eqref{eq:dimensionless-amplitude}. 
We thus find Eq.~\eqref{eq:effective-XXZ-model} by substituting it into the formula
\begin{align}
    \hat{\mathcal{H}}_{\mathrm{eff}}(A) = & \frac{\Omega}{2\pi} \int_{0}^{2\pi/\Omega} dt\; \hat{\mathcal{U}}(t) \hat{\mathcal{H}}_{0} \hat{\mathcal{U}}^{\dagger}(t).
\end{align}

Let us thereby evaluate $ \hat{\mathcal{U}}(t) \hat{\mathcal{H}}_{0} \hat{\mathcal{U}}^{\dagger}(t) $. Because of the commutation relation $ [\hat{S}_{i}^{z}, \hat{\mathcal{U}}(t)] = 0 $, we obtain
\begin{align}
    & \hat{\mathcal{U}}(t) \hat{\mathcal{H}}_{0} \hat{\mathcal{U}}^{\dagger}(t) \\
    = &
    - \frac{ J_{\perp} }{2} \sum_{\pair{i}{j}} 
        \hat{\mathcal{U}}(t)\qty(
            \hat{S}_{i}^{+}\hat{S}_{j}^{-}
            + \hat{S}_{i}^{-}\hat{S}_{j}^{+}
        )
        \hat{\mathcal{U}}^{\dagger}(t)
 - \bar{J_{\parallel}} \sum_{\pair{i}{j}} \hat{S}_{i}^{z} \hat{S}_{j}^{z}. \label{eq:expression-of-UHU}
\end{align}
Furthermore, using the formula
\begin{align}
    \mathrm{e}^{-i\hat{A}} \hat{B} \mathrm{e}^{i\hat{A}} = \sum_{n = 0}^{\infty} \frac{(-i)^{n}}{n!} \qty(\mathrm{ad}\hat{A})^{n}\hat{B},
\end{align}
where $ \mathrm{ad}\hat{A}\cdot \hat{B} \coloneqq [\hat{A}, \hat{B}] $, for any pair of operators $ \hat{A} $ and $ \hat{B} $, 
one of the terms in the parentheses in Eq.~\eqref{eq:expression-of-UHU} can be written as
\begin{align}
    & \hat{\mathcal{U}}(t)\hat{S}_{i}^{+}\hat{S}_{j}^{-}
        \hat{\mathcal{U}}^{\dagger}(t) \\
    = &
    \sum_{n = 0}^{\infty} \frac{(-i)^{n}}{n!} \qty(
        A \sin\Omega t
    )^{n} \textstyle \qty(
        \mathrm{ad}\sum_{\pair{k}{l}} \hat{S}^{z}_{k} \hat{S}_{l}^{z}
    )^{n} \qty(\hat{S}_{i}^{+} \hat{S}_{j}^{-}) \label{eq:USSU-1}\\
    = & \sum_{n = 0}^{\infty} \frac{(-i)^{n}}{n!} \qty(
        A \sin\Omega t
    )^{n} \hat{S}_{i}^{+} (\hat{Z}_{i, j})^{n} \hat{S}_{j}^{-} \label{eq:USSU-2}\\
    = & \hat{S}_{i}^{+}
    \exp\qty[
        -i A \sin \Omega t \times\hat{Z}_{i, j}
    ] 
    \hat{S}_{j}^{-}, \label{eq:USSU-3}
\end{align}
where $ \hat{Z}_{i, j} $ is defined in Eq.~\eqref{eq:staggered-magnetization-operator}.

We can prove the equality between Eqs.~\eqref{eq:USSU-1} and \eqref{eq:USSU-2} by induction as follows. If we accept
\begin{align}
    \textstyle \qty(
        \mathrm{ad}\sum_{\pair{k}{l}} \hat{S}^{z}_{k} \hat{S}_{l}^{z}
    )^{m} \qty(\hat{S}_{i}^{+} \hat{S}_{j}^{-}) = \hat{S}_{i}^{+} (\hat{Z}_{i, j})^{m} \hat{S}_{j}^{-} \label{eq:induction-1}
\end{align}
for a non-negative integer $ m $, we obtain
\begin{align}
    \textstyle &  \qty(
        \mathrm{ad}\sum_{\pair{k}{l}} \hat{S}^{z}_{k} \hat{S}_{l}^{z}
    )^{m + 1} \qty(\hat{S}_{i}^{+} \hat{S}_{j}^{-}) \\
    = &
    \sum_{\pair{k}{l}} [ \hat{S}^{z}_{k} \hat{S}_{l}^{z}, \hat{S}_{i}^{+} (\hat{Z}_{i, j})^{m} \hat{S}_{j}^{-}] \label{eq:adjoint-1}\\
    = & \textstyle 2 \times \hat{S}_{i}^{+}  \qty[
        \sum_{\pair{k}{l}}
        \qty(
            \delta_{l, i} - \delta_{l, j}
        ) \hat{S}^{z}_{k}
    ](\hat{Z}_{i, j})^{m} \hat{S}_{j}^{-} \label{eq:adjoint-2}\\
    = & \hat{S}_{i}^{+} (\hat{Z}_{i, j})^{m + 1} \hat{S}_{j}^{-}. \label{eq:adjoint-3}
\end{align}
For the equality between Eq.~\eqref{eq:adjoint-1} and Eq.~\eqref{eq:adjoint-2}, we use the symmetry of exchanging 
indices $ k $ and $ l $. For the equality between Eq.~\eqref{eq:adjoint-2} and Eq.~\eqref{eq:adjoint-3}, we use the decomposition 
of the sum given by
\begin{align}
    2 \times \sum_{\pair{k}{l}} \hat{\mathcal{O}}_{k, l} = \sum_{k} \sum_{\pair{l}{\underline{k}}} \hat{\mathcal{O}}_{k, l},
\end{align}
where $ \sum_{\pair{l}{\underline{k}}} \hat{\mathcal{O}}_{k,l} $ denotes summation of $ \hat{\mathcal{O}}_{k,l} $ for all sites $ l $ that are connected to site $ k $. Since Eq.~\eqref{eq:induction-1} trivially holds for $ m = 0 $, it holds for any non-negative integers.

In exactly the same way, we obtain
\begin{align}
    \hat{\mathcal{U}}(t)\hat{S}_{i}^{-}\hat{S}_{j}^{+}
        \hat{\mathcal{U}}^{\dagger}(t) = & \hat{S}_{i}^{-}
    \exp\qty[
        + i A \sin \Omega t \times\hat{Z}_{i, j}
    ] 
    \hat{S}_{j}^{+}.
\end{align}
    Hence, we find
\begin{widetext}
\begin{align}
    \hat{\mathcal{U}}(t) \hat{\mathcal{H}}_{0} \hat{\mathcal{U}}^{\dagger}(t) = - \frac{ J_{\perp} }{2} \sum_{\pair{i}{j}} \qty[
            \hat{S}_{i}^{+}\mathrm{e}^{-i A \sin \Omega t \times\hat{Z}_{i, j}}
        \hat{S}_{j}^{-} + (+ \leftrightarrow -)
    ] - \bar{J_{\parallel}} \sum_{\pair{i}{j}} \hat{S}_{i}^{z} \hat{S}_{j}^{z}. \label{eq:expression-of-UHU2}
\end{align}
Using the integral formula \cite{temmeSpecialFunctionsIntroduction1996}
\begin{align}
    \frac{1}{2\pi}\int_{0}^{2\pi} d\theta \; \mathrm{e}^{-ia \sin\theta} = \mathcal{J}_{0}(a), \qq{for $ a\in \mathbb{R} $}
\end{align}
in Eq.~\eqref{eq:expression-of-UHU2}, we arrive at
\begin{align}
    \hat{\mathcal{H}}_{\mathrm{eff}}(A)
    & = \frac{\Omega}{2\pi} \int_{0}^{2\pi/\Omega} dt\; \hat{\mathcal{U}}(t) \hat{\mathcal{H}}_{0} \hat{\mathcal{U}}^{\dagger}(t)\\
    & = - \frac{J_{\perp}}{2} \sum_{\pair{i}{j}} \left[
    \hat{S}_{i}^{+} \mathcal{J}_{0}(
      A \hat{Z}_{i, j}
    ) \hat{S}_{j}^{-}
    + (+ \leftrightarrow -)\right] - \bar{J}_{\parallel} \sum_{\pair{i}{j}} \hat{S}_{i}^{z} \hat{S}_{j}^{z},
\end{align}
where $ \mathcal{J}_{0}(x) $ is the zeroth-order Bessel function of the first kind.
\end{widetext}

\bibliography{ZoteroBibTeX}

\end{document}